\documentclass[twocolumn,aps,prl]{revtex4} 
\draft
\usepackage{epsfig}
\begin{document} 
\title{A Spin-Mechanical Device for Detection and Control of 
Spin Current by  Nanomechanical Torque}  
\author{P. Mohanty$^1$, G. Zolfagharkhani$^1$, S. Kettemann$^{2,3}$, P. Fulde$^3$} \address{$^1$Department of Physics, Boston University, 590 Commonwealth Avenue,
Boston, MA 02215, USA}   
\address{$^2$ 1. Institut f\"ur Theoretische Physik, Jungiusstr. 9, Universit\"at Hamburg, Hamburg 20355, Germany,}  
\address{$^3$Max-Planck Institut f\"ur Physik Komplexer Systeme, N\"othnitzer Str. 38, D-01187 Dresden, Germany}  

\begin{abstract} 
We propose a spin-mechanical device to control and detect spin currents  by
mechanical torque. Our hybrid nano-electro-mechanical device, which contains a
nanowire with a ferromagnetic-nonmagnetic interface, is designed to measure or
induce spin polarized currents. 
Since spin carries angular momentum, a spin flip or spin transfer process
involves a change in angular momentum--and hence, a torque---which enables
mechanical measurement of spin flips. Conversely, an applied torque can
result in  spin polarization and  spin current. 
 
\vskip 0.2 in  
\end{abstract}  
\maketitle 

\section{Introduction}  
Electronics and information processing with spins promise to be anything but   
conventional \cite{spintronics-book,wolf}. In spin-based electronics, 
information is injected, stored, transferred, or manipulated with the spin 
degree of freedom. The realization of its importance---and the ensuing 
excitement---stems from the fact that successful incorporation of spins into 
conventional semiconductor technology  
will allow to combine both information storage and information manipulation 
within a single platform. Recent discovery of a number of spin-based phenomena 
such as giant magnetoresistance \cite{gmr} has also marked the beginning of a 
new era with spintronics \cite{spintronics-book,wolf}, and spin-based quantum 
information processing \cite{loss}.   
 The transport properties of d--electrons  
 in ferromagnetic transition metals like Fe, Co, Ni, and their 
 contribution  to the ferromagnetism are also 
 of important fundamental interest
\cite{aharoni,fulde2,manyala,saxena,moriya}.

At the heart of spintronics is detection and control
 of electron spins moving   
through a ferromagnetic-metallic or ferromagnetic-semiconductor 
heterostructure. Towards this end, there have been a large number of important 
experiments \cite{spin-expt} and theoretical works \cite{datta}, leading 
to a whole new dictionary for spin-based electronics---spin diode, spin 
transistor, spin pump, spin battery and spin 
filter \cite{spintronics-book,wolf}.  
Here, we propose a spin-mechanical device for detection and control of 
spins in hybrid magnetic-nonmagnetic structures by mechanical torque.   
 
Spin transport and nanomechanics form the underlying basis of the proposed 
spin-mechanical device, which is capable of detecting,
 controlling and creating 
both spin transport and spin population. The central concept of its function 
is rather simple: spin carries angular momentum and a change in the angular 
momentum due to spin transport creates a torque, which can be detected by 
a nanomechanical torsion oscillator.  
 
The deceptive simplicity of the idea contrasts its rich 
history, both in theory and experiment. The original proposal 
to detect changes in the magnetization by measuring the associated torque 
dates almost a century back to the work of Richardson \cite{richardson},  
and Einstein and de Haas \cite{einstein}.
 The reverse effect, inducing magnetization by rotation
 has been  suggested and measured
 soon    by  Barnett \cite{barnett}. Another  mechanism  
is the associated Wiedemann effect \cite{wiedemann}:
 a current flowing through  
a ferromagnet, such as  a Ni-- or Fe wire, in a parallel magnetic field  
induces a torque on the wire due to magnetoelastic interactions. 
 
Motivated by these historical experiments it was recently pointed out that 
a current flowing through a ferromagnetic-nonmagnetic (FM-NM) interface produces 
a mechanical  torque. In a ferromagnet, a current has 
an associated spin current, which is absent in a nonmagnetic metal. 
Therefore, there is a sink or source of angular momentum at an FM-NM interface,  
depending on the current direction,
 which results in a torque\cite{berger},  which 
 can be transfered into  a mechanical torque of the device \cite{fulde}. 
This is precisely the underlying principle of the classic optical 
 experiment by Beth 
\cite{beth} in which the change in the angular momentum of circularly  
polarized photons through an optically active glass is measured; the torque 
on the glass plate results due to the change in polarization at the interface. 
 This allowed for the first  time to prove experimentally 
    the quantization of the  spin of photons. 

 A measurement of the torque at an FM-NM interface
 would allow for a
 determination of the relative contributions
 of different types of electrons to
 the current, e.g., of s and d electrons
 in the case of Fe \cite{fulde2,aharoni}. Those electrons are
 polarized to a different degree because of
 their different bandwidths.
 From an experimental measurement of 
 the tunneling density of states, magnetic 
 polarisations of $\alpha= .11,.34,.44$
 for Ni, Co and Fe have been estimated, while a theoretical calculation for s-electrons alone would give for Ni, $\alpha_s=.01$\cite{fulde2}. 
 Until
 now, there is no direct measurement available of
 the  relative contribution of the  s and d electrons to
 the current.

\section{Spin Flip Torsion Balance}
 It is the purpose of this article to 
  show that this spin flip torsion balance effect 
 can be strongly enhanced in a nanomechanical device  when  
  a nanowire containing a FM-NM interface is grown on top of a suspended 
nano-electro-mechanical structure (NEMS). Thereby, 
 the  spin transport can be 
measured by the induced mechanical torque  in the NEMS torsion device.  
 The inverse is also true: a torque at the FM-NM 
interface produces a potential difference between the two metals\cite{barnett,landau}. 
Accordingly, a spin current can be generated by applying an external torque. 
In this paper, we perform a detailed analysis to show that  
such a device can be used to detect, control and induce spin current. 
 
In a nanowire with a ferromagnetic (FM) and a nonmagnetic (NM) 
part, the ferromagnet is uniformally spin  polarized in a direction 
parallel to the wire. 
Application of a magnetic field parallel to the wire further elongates  
the magnetization. When a current $I$ is passed through the wire, 
a fraction $\alpha/2$ of the electrons passing through the interface between 
the FM and NM flip their spin. Here,  $\alpha$ is the degree of 
magnetic polarization of the electrons  
contributing to the current in the ferromagnet.  
\begin{figure}[t] 
\epsfxsize=8.0 cm 
\epsfysize=6.0 cm 
\epsfbox{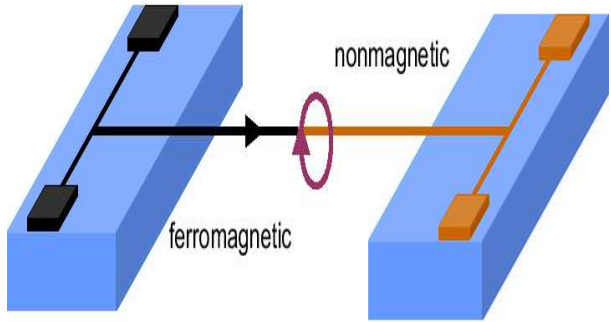} 
\caption{(Color online) Schematic diagram, showing    
 a suspended ferromagnetic-nonmagnetic (FM-NM) wire.   
 The black arrow denotes the direction of the magnetisation.
A spin-transfer process at or beyond the FM-NM   
interface torts the wire as indicated.}  
\end{figure}  
 
Each spin flip results in a change in the angular momentum $\Delta L = \hbar$.  
The number of electrons $\Delta N$ passing the interface in a time  
interval $\Delta t$ is $\Delta N = I \Delta t/e$, where $I$ is the electrical 
(charge) current, and $e$ the electron charge. The associated change
in
 angular 
momentum is $\Delta L/\Delta t = (\hbar \alpha/2) I/e$, which results in a torque   
\begin{equation} \label{dc} 
{\vec{T}}_{\rm spin} = {\hbar \over 2} \alpha {I \over e}  \hat{x}~~~. 
\end{equation}  
 Here, $\hat{x}$ is the unit vector along the wire axis. 
The higher the spin polarization of the conducting electrons, the larger is the 
resulting torque. 
 Thus, that torque is a direct measure of the spin-polarisation of the 
 itinerant electrons in the ferromagnet. 
If the current increases, the resulting torque increases linearly.   
 
There is an additional contribution {\bf T}$_{\rm W}$  
to the net torque because of the Wiedemann effect. This contribution is orders of 
magnitude smaller than ${\vec{T}}_{\rm spin}$ in nanomechanical systems as we show below. 
{\bf T}$_{\rm W}$  originates from  
the circular magnetic field $H_{\rm c}= I/2\pi R$ perpendicular to the magnetization  
axis of the FM part of the nanowire in presence of a current I. It leads to small changes  
in the direction and the absolute value of the magnetization by the magneto-elastic 
interaction. The resulting torque {\bf T}$_{\rm W}$ has an associated  
torsion angle $ \theta_W = (L \lambda/R)\delta M_c/M_0$.  
Here, the radius of the wire is $R$, its length $L$,  and $M_0$ 
is the magnetization along the wire axis.  $\lambda = \Delta \ell/\ell$ is the 
magnetostrictive coefficient which characterizes the relative length changes 
$\Delta \ell/\ell$ due to the changes of the magnetization $|\delta {\bf M}|$.  
Here, $\delta M_c = \chi_c H_c$  
where $H_c $ is the field generated by the current $I$ 
and $\chi_p$ is   the magnetic susceptibility 
 to the  magnetic field $H_c$,  pointing circularly   around the 
  wire axis, when 
 the single domain ferromagnet is  polarised along the wire.  
The torque is then given by $ T_W = K \theta_W$,  
where for a wire, $K = \frac{\pi}{2} G R^4/L$ and  $G$ is the 
shear modulus of the oscillator.   
The ratio between T$_{\rm W}$ and T$_{\rm spin}$ is   
\begin{equation} 
\frac{T_{\rm W}}{T_{\rm spin}} = \frac{R^2 Ge \lambda    \chi}{2 \hbar \alpha 
M_0}~.  
\end{equation}  
The magnetostriction  coefficient $\lambda$ 
is in general quite small (relative to the unpolarised state, it is on the order of $10^{-6}$) and  
both its sign and magnitude depend 
on the material and the material growth direction. 
 Here,  we are interested in the magnetostriction as  caused by  the reorientation of the magnetisation  
 due to  the circular field,    relative to the state when the
 wire is fully  polarised along the wire.    
In that case, $\lambda$  is a function of the magnetization $M_0$ and 
applied magnetic field $H_c$, and it displays strong hysteresis\cite{lee}.    
From magnetoelastic theory one expects   
$\lambda \approx  \Lambda  (\delta M/ M_0)^2$\cite{lee}, although the exact 
functional form may depend on the material and crystal orientation.  
 
From  magnetostriction measurements on Fe or Ni \cite{lee} 
we estimate  $\Lambda \approx  10^{-5}$, when the ferromagnet is  
magnetized along the wire with typical magnetization  $M_0 =\frac{1}{4 \pi}  10^6  A/m$.
  We find  
with $G = 10^{10} N/m^2$,  
\begin{equation}  
\frac{T_{\rm spin}}{T_{\rm W}} = \alpha  8.5 ~ 10^{-5}/(\rho^4 \chi_p^3 {\bf j}^2), 
\end{equation} 
where $ R = \rho (mm)$, and $j =  {\bf j} (A/mm^2)$ is the current density.  
In the classical macroscopic experiments\cite{wiedemann} 
with $\rho \approx  0.1$  and ${\bf j} =1$, 
the Wiedemann effect and the spin flip torsion effect 
can be  of the same order of  magnitude.  
However, in a nanomechanical torsion oscillator,  
the wire can be as thin  as $\rho = 10^{-4}$,  
while the current density can still be  ${\bf j} =1$, the spin flip torsion balance  
can exceed the Wiedemann effect by orders of magnitude. 
 When applying the circular magnetic field, the single magnetic domain, magnetised along the wire 
 has  some rigidity so that the  
  corresponding  magnetic susceptibility $\chi_c$ is small.
 Taking
 for a ferromagnetic cylinder  polarised along its axis  
 a typical value  $\chi_c =1$\cite{hernando}
  one finds $T_{\rm spin}/T_W \approx 10^{12}$.
 Although the circular magnetic susceptibility can depend on the growth direction of the ferromagnetic film, 
 and can change with its dimension, we can conclude that the Wiedemann
 effect is negligible for any theoretically possible  value of the 
 circular  magnetic susceptibility, 
 ranging  between the Pauli susceptibility $\chi_P \sim 10^{-4}$, and 
 values for macroscopic ferromagnets $\chi \sim 10^3$.  
 
\noindent{\it Spin-Mechanical Balance:}   
The spin-transfer torque can be measured with a suspended NEMS torsion balance 
to which the half-ferromagnetic, half-nonmagnetic nanowire is rigidly 
attached, as shown in Fig. 1.  The magnitude of the torque, 
Eq.(\ref{dc}) along the wire axis is given by  
\begin{equation} 
T_{\rm spin} =  3.3 ~~ 10^{-19} \alpha  { \bf I} N m, 
\end{equation} 
where ${\bf I}$ is the current in units of  $mA$.  Thus, 
 the torque is within the 
sensitivity range of existing NEMS oscillators\cite{cleland,dissipation}.
 The torque caused by the finite 
spin flip rate, Eq. (\ref{dc}), is in balance with the torque due to the 
elastic torsion  of the wire, as well as a torque due to the finite inertia of 
the wire, and a  friction torque when the torsion angle changes in time. A 
ferromagnetic wire, on top of the NEMS oscillator, undergoing torsion by an 
angle $\theta$ results in a torque $T_{elastic} = K \theta$. $L$ is 
the length of the ferromagnetic part of the wire. If the angle 
$\theta$ changes with time, it will result in an inertial torque $T_{inertia}$ 
due to the moment of inertia $J$ about the nanowire axis: $T_{inertia} = 
Jd^2\theta/dt^2$.   
The equation of motion for the torsion angle $\theta$ is 
\begin{equation} \label{ldeq} 
J{d^2\theta \over dt^2} + \gamma {d\theta \over dt} + K \theta = T_{\rm spin} + T_W.  
\label{damped}  
\end{equation}  
where $\gamma d\theta/dt$ describes frictional damping,   
and $K=  {\pi G \over 2 L} R^4$ for a wire of radius $R$.  
\noindent  
Let us first consider the case when $I$ is large and $T_{\rm spin}$
assumes the form Eq. (\ref{dc}). (Since  the torque due to the
Wiedemann effect $T_W$, has been shown to be small above, we will 
 disregard that term in the following.)  
 
\noindent  
{\it Case I: Constant Spin Current (Equilibrium Torsion):}  
If the spin polarizing current through the nanowire is time independent,   
then $T_{\rm spin} =\mbox{constant}$, which will result in an equilibrium torsion  
angle $\theta_0$:  
\begin{equation}  \label{thetaconst} 
\theta_0 = {\hbar \over 2 K} {I_0 \over e} \alpha.  
\end{equation}  
\noindent  
This simple expression shows that $\theta_0$ is linearly proportional to the 
 current $I_0$, and the spin polarization $\alpha$. We note that the torsion angle 
 can be strongly enhanced by reducing the radius  of the wire $R$, since $K \sim 
R^4$.  
\noindent  
{\it Case II: Spin Current at a Finite Frequency $\omega$:}  
A finite-frequency driving current $I_0 cos\omega t$ results in a torque, 
$T_{\rm spin} = (\hbar \alpha/2e) I_0 cos\omega t$. With this expression on the right 
hand side of Eq.(\ref{damped}), the analysis is  that of a driven 
damped oscillator. The solution for the torsion angle is  
\begin{eqnarray}  
\theta(\omega, t) = {\hbar I_0 \alpha/2e \over J[(\omega_0^2-\omega^2)^2   
+ 4\omega^4/Q^2]^{1/2}}cos(\omega t +\phi), \nonumber \\   
\phi = \arccos \left[ (\omega_0^2 - \omega^2 )/\sqrt{(\omega_0^2-\omega^2 )^2 + 
4 \omega_0^2 \omega^2/Q^2} \right].  
\end{eqnarray} 
The quality factor of the structure $Q$ is 
defined through the characteristic exponential decay of the torsion angle in 
the absence of driving torque, $\theta = \theta_0 cos(\omega 
t+\phi)e^{-\Gamma t}$ so that $Q = \omega_0/\Gamma = 2J\omega /\gamma$. 
 
On resonance $\omega = \omega_0$, the resulting torsion angle from the spin 
transfer torque reaches its maximum value:   
\begin{equation}  
\theta_{max} = \hbar {Q \over 4J\omega_0^2} {I_0 \over e} \alpha.  
\end{equation}  
It is clear from this expression that the signal is maximized for optimal 
mechanical parameters: high quality factor $Q$ and low resonance frequency 
$\omega_0$.   In order to be measurable, this torsion angle  
has to be larger than the thermal fluctuations, $\delta \theta$. 
By setting the elastic energy $E(\delta \theta)$ equal to  
$ k_B T/2$ (equipartition), one obtains  
\begin{equation} 
E(\delta \alpha ) = \int_0^{\delta \theta} d \theta D_{elastic} ( \theta ) =   
 \frac{\pi}{4}  (\delta \theta)^2  G \frac{R^4}{L}, 
\end{equation} 
 so the torsion angle may maximally fluctuate by  
\begin{equation} \label{thermal}
\delta \theta = \sqrt{\frac{2}{\pi}} \frac{1}{R^2} \sqrt{ \frac{k_B T L}{G} }.   
\end{equation} 
	 
Up to this point, the torsion angle is found not to depend on the spin 
flip rate $1/\tau_s$ explicitly. Let us consider the situation when the 
current is reduced such that the transfer rate $I/q$ becomes 
smaller than  $1/\tau_s$, and the spin flips occur one by one.

\section{Random Spin Flip}  
For low but constant driving currents the electron transfer rate at the 
interface can be smaller than the spin-flip scattering rate 1/$\tau_s$. In this 
case, the spin transfer torque can be considered to occur randomly at times, 
$t_l$, with the average  amplitude, 
\begin{equation} 
T_{\rm R} =   \hbar \hspace{.1cm} \frac{\alpha}{2}  \frac{1}{\tau_s} = 
\frac{\alpha}{2} ~ 1.1~ 10^{-25} 
N m  \frac{1}{{\bf \tau_s}}, 
\end{equation}   
where $\tau_s = {\bf  \tau_s} (ns)$.  
Thus, the time dependent driving  torque can be modeled by  
\begin{equation} 
T_{\rm spin} ( t ) = T_{\rm R}  \sum_l \left[ \Theta ( t - t_l ) - \Theta (t - 
\tau_s - t_l) \right],  
\end{equation} 
where $\Theta(x)$ is the step function,  $\Theta(x) = 1$ for $x > 1 $,  
and $\Theta(x) = 0$, otherwise.   
The equation of motion  Eq.(\ref{damped}) with the right hand side,   
$T_{s} (t) $, 
can be solved by Fourier transform:  
\begin{eqnarray} 
\theta (t) &=& \int \frac{d \omega}{2 \pi} \frac{T_{\rm R} }{\delta + i \omega 
}\frac{1}{K - \omega^2 J + i 
\gamma \omega} 
\nonumber \\  && 
 \sum_l \left[ \exp (i \omega (t -  t_l)) - \exp (i \omega (t - \tau_s - t_l))   
\right],   
\end{eqnarray} 
where $ \delta \rightarrow 0^+$.   
Performing the residuum integral, we obtain, 
for $\omega_0 < 1/\tau_s$; the time dependence of the torsion angle  
reduces to a sum of sinusoidal oscillators with phase shifts at   
random times $t_l$, when a single spin flip occurs.   
\begin{eqnarray} \label{thetarandom} 
\theta (t) &=& \frac{\hbar \alpha}{ 2 J \omega_0 \sqrt{1 - 1/Q^2}} \sum_{l, t_l < t} 
\exp \left( - (t-t_l) \frac{\omega_0}{Q} \right)  
\nonumber \\ &&  
\sin \left( (t-t_l) \omega_0 \sqrt{1 - 1/Q^2} \right).   
\end{eqnarray} 
While the amplitude of the torque does depend on the  
spin flip rate $1/\tau_s$, the torsion angle does not depend  
on it, and is strictly a function of the current $I$.  
For $I/e \gg \omega_0$ it can be shown that Eq. (\ref{thetarandom}) 
reduces to Eq. (\ref{thetaconst}). 
 
 When the  resonance frequency of the oscillator exceeds the  
spin flip rate, $ \omega_0 > 1/\tau_s$, one obtains  for large quality factor $Q \gg 1$, 
the simplified expression,  
\begin{eqnarray} 
\theta(t) &=& \theta_0  + \theta_{\rm R} \sum_{t_l < t} \cos \left[ ( t - t_l ) 
\omega_0 \right] 
\nonumber \\  
& -&  \theta_{\rm R}  \sum _{t_l + \tau_s < t}  
\cos \left[ ( t - t_l -\tau_s ) \omega_0 \right].  
\end{eqnarray}   
Here $  \theta_{\rm R} = T_{\rm R}/K$.  
Thus, when   $ \omega_0 > 1/\tau_s$, both the torque and the amplitude of the  
time dependent torsion angle $\theta_{R}$ do depend directly  
on the spin flip rate $1/\tau_s$.  
 
The nature of the random spin-flip torque is not known a priori. It depends 
on the microscopic physical mechanism of spin relaxation via a number of 
channels such as coupling to local magnetic moments, nuclear magnetic moments,  
and dominantly, via spin--orbit interaction to phonons by the Elliot-Yafet 
mechanism\cite{elliot,yafet,das-sarma}. The spin-orbit interaction modified by phonons 
can break momentum conservation in a periodic crystal 
\cite{overhauser,grimaldi} and hence result in relaxation. 
Furthermore, as shown by D'yanokov and Perel\cite{perel}, in crystals 
lacking momentum-inversion symmetry, the lifting of spin degeneracy  by  
spin-orbit scattering\cite{kittel},  operates as a momentum-dependent internal magnetic 
field, and contributes to spin relaxation.  
Temperature dependence of the spin relaxation rate $1/T_1$ in the 
Elliot-Yafet mechanism is the same as the temperature dependence of 
resistivity: (i) $1/T_1 \sim T$ for $T > T_{Debye}$, (ii) $1/T_1 \sim T^5$ (at 
low temperature). D'yanokov-Perel mechanism results in a spin 
relaxation rate proportional to the momentum relaxation time.  

In the nonlinear regime, the resonance of the oscillator in the presence of  
random torques can be enhanced by nonlinear self-excited oscillations. This effect 
is due to the nonlinear dependence of the torque on the angle $\theta$ \cite{bow}.  
In fact, Eq. (\ref{ldeq}) becomes nonlinear with the magnetization dynamics.

\section{Spin Current Generation}
Applying a torque,  a  potential difference along the wire axis is  
created\cite{landau},  as given by 
$e V(t)  = - (2 m/e g)  Vol. M_z  {d \theta/d t}$.
 With proper design of the device this potential can ve used to drive   a time dependent current with density 
\begin{equation}  \label{barnett} 
 j_B (t)  = - \sigma  \pi R^2 \frac{2 m}{e^2 g} M_z \frac{d \theta}{d t}.  
\end{equation} 
where $\sigma$ is the conductivity, and $g$ is the gyromagnetic coefficient. 
 Writing  $M_z = {\bf M_z} (10^6/ 4 \pi) A/m$, where the dimensionless
 ${\bf M_z}$  is typically on the order of one, and $ R = \rho \mu m$, 
 we obtain inserting as the typical metallic  conductivity 
$\sigma = 10^7/\Omega m$, 
 $  j_B  \sim - 10^4 \rho^2 {\bf M_z} (d\theta/d t)  s A/cm^2$. 

Another source of potential difference is from magnetostriction: when  
the magnetized wire is torted, it results in a circular component.  
This so-called Matteucci effect induces a potential difference, 
and, hence, in a closed circuit
 a current density\cite{matteucci}:  
 \begin{equation} 
j_M = - 2 \sigma \frac{R^2}{L} \lambda G \frac{\chi}{M_z} \frac{d \theta}{d t}.  
\end{equation}  
 Inserting the typical values of $G$, $\lambda$, and $\sigma$,
 as given above, 
 we obtain $j_M \sim - 10^{-3} \rho^2/{\bf L}   (d\theta/d t)  s A/cm^2$, 
 where $L = {\bf L} \mu m$, which is thus negligible compared to the 
 current due to the Barnett effect, Eq. (\ref{barnett}).


\begin{figure}[t] 
\epsfxsize=8.0 cm 
\epsfysize=6.0 cm 
\epsfbox{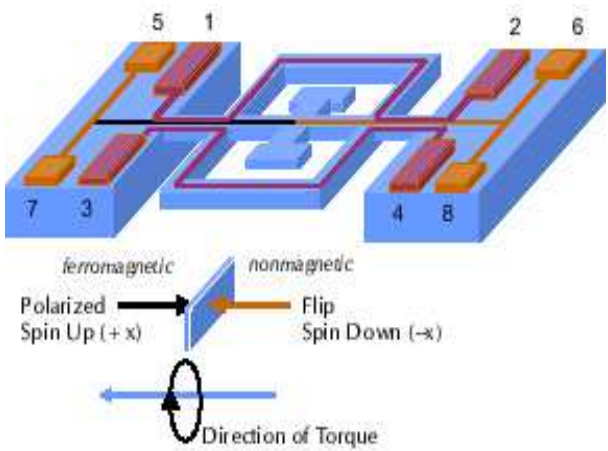} 
\caption{(Color online) Schematic diagram of the proposed device.   
It contains a ferromagnetic-nonmagnetic (FM-NM) wire   
on top of a suspended two-element torsion oscillator.   
A spin-transfer process at or beyond the FM-NM   
interface causes a mechanical torque,   
which twists the suspended structure since the FM-NM wire is rigidly   
attached to the torsion oscillator. Because of small size, this  
device has extremely high torque sensitivity at low temperatures.   
Conversely, a spin polarization can be induced by applying a torque  
to the outer paddle, which will result in an imbalance  
in the spin states. The polarity of the spin population along the  
nanowire is governed by the conservation of angular momentum, which  
enables the device to operate as a spin battery.}  
\end{figure}  
\section{Experimental Realisation}
Sensitive measurement of nanomechanical torque generated due to spin transfer
in the FM-NM nanowire can be done in a number of configurations. We discuss
a specific device which allows ultra-high detection sensitivity. Figure 2 shows
the schematic diagram of such a hybrid device, which contains the 
ferromagnetic-nonmagnetic nanowire on top of a suspended two-element torsion
oscillator. Since the nanowire is fabricated on top of the torsion oscillator,
the torque generated in the nanowire will translate to a torque in the entire
structure, including the torsion oscillator, modifying accordingly the
mechanical parameters, the  inertia  $J$, the bulk modulus $G$ and the
resonance frequencies of the device  entering Eq. (\ref{damped}). 

The hybrid structure contains three sets of electrical components. The first one
(5-6-7-8) is the central FM-NM nanowire with a typical thickness of 40 nm -- 100 nm and
lithographically-defined width of 60 nm -- 100 nm. The other two electrical
components (1-2 and 3-4), the two electrodes symmetrically placed on both sides of the
large outer paddles, are designed to allow both detection and control of the
torque\cite{dissipation}, and hence of the spin transfer. In presence of a magnetic field, applied in
such a way that the torsion of the outer paddles will enclose a finite area, the
torsion of the structure will induce an electromotive voltage on the electrode (1-2)
by the Faraday effect. Likewise, an applied current through the other electrode (3-4)
can generate a control force $F_{drive} = I_{control}LB$ or torque on the structure.
This control torque can be used, for example, to induce spin polarization in the
structure. 

Measurement of the spin-induced torque and application of the control torque 
in the proposed device can be performed by the magnetomotive technique \cite{dissipation}.  
Elsewhere, we have discussed in detail low temperature measurements with
torsion oscillators similar to the one schematically shown in Figure 2 \cite{dissipation}.  
Because of low temperature and small size, the proposed device can  
operate with unprecedented force sensitivity of $10^{-16} N/\sqrt{Hz}$ or a corresponding 
torque sensitivity of $10^{-21} N.m/\sqrt{Hz}$, as recent experiments on similar 
devices have demonstrated \cite{smallforce}. Correspondingly, the minimum detectable force 
or torque can be obtained by using the intrinsic bandwidth of the device $\omega/2\pi Q$.  
For a similar control device, we have already obtained a minimum detectable force of $48 \times 10^{-18} N$ or a minimum detectable torque of $48 \times 10^{-23} N.m$ for a moment arm of $\sim 10 \mu m$ at a temperature of 4 K \cite{smallforce}. Elsewhere \cite{dissipation}, we have extensively 
characterized a set of control torsion devices and their dependence on temperature, field, size, 
and frequency. 

Experimental realization of the proposed device relies on the nanofabrication 
of a multilayer nanomechanical structure. Recently, we have successfully fabricated 
a set of suspended devices of single-crystal silicon with a half gold, half cobalt nanowire 
designed to resonate in the range of 1-10 MHz. For  device parameters of $L = 10 \mu m$, 
$R = 100 nm$, $\omega_0/2\pi \simeq 1 MHz$, and $Q \simeq 10^4$ (at 300 mK), the maximum  
torsion angle due to the spin torque is $\simeq 4.0 \times 10^{-2}$ radian for a drive 
current of 10 nA. For comparison, the rms value of the fluctuations in the torsion angle due to  
thermal noise, Eq. (\ref{thermal}),
 in the same device is $\sim 2 \times 10^{-5}$ radian at 300 mK.     
 To observe the  maximal torsion,
 the magnetisation of the Cobalt nanowire is required to be
 oriented  parallell
 to the wire axis. Indeed, it has been shown experimentally, 
 with the magneto optical Kerr effect,  that the easy axis of Cobalt thin films is inplane for a 
 film thickness exceeding $d = 1nm$. Below that a reorientation 
 perpendicular to the film has been observed\cite{inplane}, and explained due to 
 the crystal and interface anisotropy, which overcomes the shape anisotropy 
  when the film contains only  few monolayers\cite{reorientation}. 
  Ideally, one should have a single domain along the wire, 
 which is in the chosen wire geometry the one favoured by the 
 shape anisotropy. If one has domains with alternating magnetisation
 along the wire, there will be excitation of 
 torsional modes with  wave lengths on the 
 order of the domain size, in addition to the zero mode, considered
 above.  In order to maximize the torsion signal,  
 one can obtain a single  domain 
 by applying  a weak magnetic field 
   of  not more than  100 Gauss  along
 the  wire\cite{inplane}.

\section{Conclusions} 
In summary,  we propose a novel nanomechanical device, a sensor  
of spin dynamics at the ferromagnetic--nonmagnetic  
interface of a wire fabricated on top of a suspended torsion oscillator. We 
explicitly derive closed-form expressions for the 
torque created by spin currents, 
and other physical mechanisms. This elementary device can be used to detect,  
control and induce spin currents 
by selectively applying and measuring the torque in the nanomechanical 
resonator. The basic structure can be further modified to create  
devices for eventual use in spintronics, and spin information processing. 
 
\section{Acknowledgements}

We acknowledge helpful discussions with G. Bergmann, C. Chamon and D. Loss, 
  K. H. M\"uller, and E. Vedmedenko. 
The work in Boston University is supported by National Science
Foundation (DMR-0346707 and ECS-0210752) and Sloan Foundation.
 S. K. acknowledges the hospitality of Boston University and  of the MPI
for Physics 
of Complex Systems in Dresden,  and
 support  by DFG, SFB 508, A9. 
 
%
%
  

\end{document}